\PassOptionsToPackage{numbers,sort&compress}{natbib}
\documentclass[sigconf]{acmart}
\usepackage{balance} 
\usepackage{natbib} 
\usepackage[skins]{tcolorbox}
\usepackage{multirow}
\usepackage{url}
\usepackage{listings}

\usepackage{appendix}

\usepackage{amsfonts}
\usepackage{amsmath}
\usepackage{algorithmic}
\usepackage{graphicx}
\usepackage{textcomp}
\usepackage{xcolor}
\usepackage{threeparttable}
\def\BibTeX{{\rm B\kern-.05em{\sc i\kern-.025em b}\kern-.08em
    T\kern-.1667em\lower.7ex\hbox{E}\kern-.125emX}}
    
\usepackage{color}
\usepackage{setspace}
\usepackage{float}
\usepackage{array}

\definecolor{dkgreen}{rgb}{0,0.6,0}
\definecolor{gray}{rgb}{0.5,0.5,0.5}
\definecolor{mauve}{rgb}{0.58,0,0.82}

\tcbset{
  my box/.style={
    enhanced,
    colframe=#1!80,
    colback=#1!10,
    attach boxed title to top left={xshift=0.2cm, yshift=-0.2cm},
    boxed title style={
      colback=#1!80,
      outer arc=0pt,
      arc=0pt,
      top=0pt,
      bottom=0pt,
    },
  },
}

\newtcolorbox{result-rq}[1]{
  my box=black,
  title=#1,
  boxrule=1.2pt,top=3pt,bottom=1.5pt,left=6pt,right=6pt
}

\AtBeginDocument{%
  \providecommand\BibTeX{{%
    Bib\TeX}}}

\setcopyright{acmlicensed}
\copyrightyear{2025}
\acmYear{2025}
\acmDOI{XXXXXXX.XXXXXXX}
\acmConference[Internetware 2025]{the 16th Asia-Pacific Symposium on Internetware}{June 20-22, 2025}{Trondheim, Norway}
\acmISBN{978-1-4503-XXXX-X/2018/06}

\begin{document}

\title{Issue Retrieval and Verification Enhanced Supplementary Code Comment Generation}



\settopmatter{authorsperrow=4}

\author{Yanzhen Zou}
\email{zouyz@pku.edu.cn}
\affiliation{%
    \institution{Key Lab of HCST(PKU),\\ MOE; SCS,\\ Peking University}
    \country{China}
}

\author{Xianlin Zhao}
\email{zhaoxianlin@pku.edu.cn}
\affiliation{%
    \institution{Key Lab of HCST(PKU),\\ MOE; SCS,\\ Peking University}
    \country{China}
}

\author{Xinglu Pan}
\email{bjdxpxl@pku.edu.cn}
\affiliation{%
    \institution{Key Lab of HCST(PKU),\\ MOE; SCS,\\ Peking University}
    \country{China}
}

\author{Bing Xie}
\authornote{Bing Xie is the corresponding author.}
\email{xiebing@pku.edu.cn}
\affiliation{%
    \institution{Key Lab of HCST(PKU),\\ MOE; SCS,\\ Peking University}
    \country{China}
}

\begin{abstract}
Issue reports have been recognized to contain rich information for retrieval-augmented code comment generation. However, how to minimize hallucinations in the generated comments remains significant challenges.
In this paper, we propose \textbf{IsComment}, an issue-based LLM retrieval and verification approach for generating method's design rationale, usage directives, and so on as supplementary code comments.
We first identify five main types of code supplementary information that issue reports can provide through code-comment-issue analysis. 
Next, we retrieve issue sentences containing these types of supplementary information and generate  candidate code comments. 
To reduce hallucinations, we filter out those candidate comments that are irrelevant to the code or unverifiable by the issue report, making the code comment generation results more reliable.
Our experiments indicate that compared with LLMs, IsComment increases the coverage of manual supplementary comments from 33.6\% to 72.2\% for ChatGPT, from 35.8\% to 88.4\% for GPT-4o, and from 35.0\% to 86.2\% for DeepSeek-V3.
Compared with existing work, IsComment can generate richer and more useful supplementary code comments for programming understanding, which is quantitatively evaluated through the MESIA metric on both methods with and without manual code comments.
\end{abstract}



\keywords{Supplementary Code Comment, Issue Report, Large Language Model, Verifiable Text Generation}


\maketitle

\section{Introduction}

The supplementary nature of code comments has gained much attention in recent studies~\cite{MESIA,10.1145/3510003.3510152,PhilosophyOfSoftware,10.1145/1085313.1085331,8719434}. 
When comprehending and reusing a method, developers often expect its code comments can provide more supplementary information beyond the method's signature or code itself \cite{MESIA}, such as design rationales and usage directives. 
However, unlike generic comments~\cite{10.1145/3510003.3510152} that often simply repeating the code, these supplementary code comments are notably limited in application, which motivated the research on supplementary code comment generation.

\begin{figure*}[t]
    \centering
      \includegraphics[width=1.0\linewidth,]{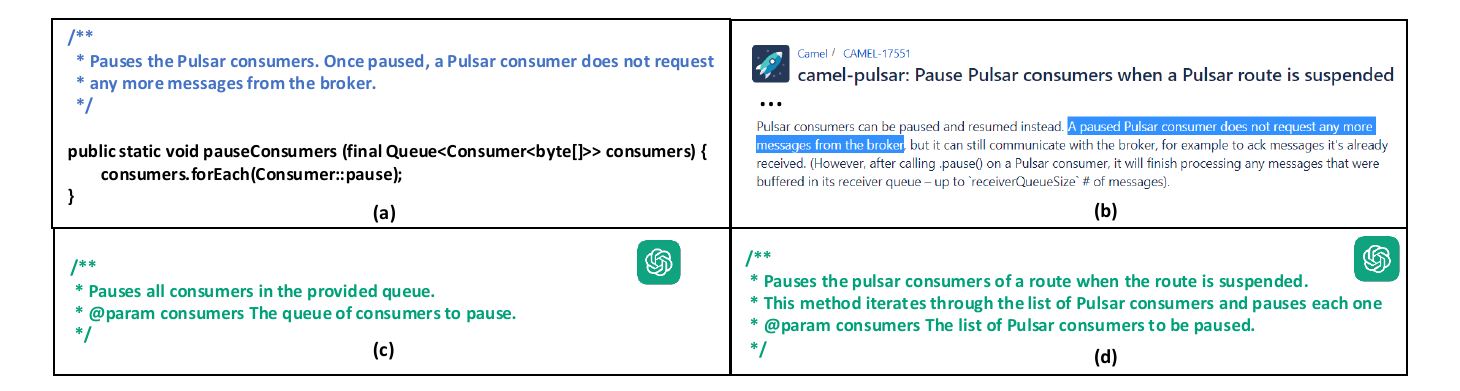}
      \vspace{-5mm}
    \caption{The comments generated by the large language model ChatGPT using code and related issue report.}
      \label{fig:Example}
      \vspace{-3mm}
\end{figure*}

Fig~\ref{fig:Example} (a) illustrates a supplementary comment for the method \emph{``pauseConsumers''}~\footnote{This method is from the Apache Camel project~\cite{Camel}.}, stating: ``\emph{Once paused, a Pulsar consumer does not request any more messages from the broker}''. This comment provides additional information about the code method's implication, which is crucial to understand the code method. Compared with the summarizing comment ``\emph{Pauses the Pulsar consumer}'', supplementary comments offer more insightful information for code understanding and maintenance.



Existing code comment generation approaches~\cite{iyer2016summarizing,hu2018deep,ahmad2020transformer,10.1145/3238147.3238206,9284039,10.1145/3324884.3416578,9678882,9678724,leclair2019neural,icsm/LuL24}, including those based on large language models (LLMs)~\cite{10.1145/3551349.3559548,2023arXiv230512865S,10.1145/3551349.3559555,10.1145/3597503.3608134,10.1145/3597503.3639183,coling/CaoHHW025,emnlp/LomshakovPSBLN24}, struggle to generate supplementary comments because they often cannot be  derived from the code.
Various studies~\cite{2023arXiv230909558P,MESIA} have reported that LLM-generated comments can sometimes even surpass manual reference comments in existing benchmark datasets. 
However, for the \emph{``pauseConsumers''} method shown in Fig~\ref{fig:Example} (a), when we prompt the LLM ChatGPT with instructions like ``Generate code comments for the code: \{code\}", as illustrated in Fig~\ref{fig:Example} (c), the generated comments primarily describe the literal meaning~\cite{10054429} of the method, failing to convey the essential information about its implication.



Considering issue reports typically discuss various development  problems and contain rich information for program understanding~\cite{Arya2019AnalysisAD,9678864}, we try to leverage Issue and Retrieval-Augmented Generation (RAG)~\cite{2023arXiv231210997G} to enhance LLMs. As shown in Fig~\ref{fig:Example} (b), the \emph{``pauseConsumers''} method is associated with an issue report (ID CAMEL-17551). A sentence in the issue report clearly says \emph{``A paused Pulsar consumer does not request any more messages from the broker''}, which is the desired information in the comment of the code method. 
However, as illustrated in Fig~\ref{fig:Example} (d), simply adding the entire miscellaneous issue to the prompt does not yield the desired supplementary comment. 
Therefore, it is necessary to first retrieve the relevant information from the issue report and then prompt LLMs to generate the supplementary comment more effectively.

Based on the analysis and observations above, we identify three challenges in leveraging issue reports and LLMs to generate supplementary code comments.
First, the information in issue reports is diverse, and further research is needed to identify the types of supplementary information that they can provide for code comments.
Second, the retriever often becomes a bottleneck when dealing with the diverse types of code supplementary information in issue reports. For example, the retriever may retrieve irrelevant content or miss the desired code supplementary information, which complicates the generation of supplementary code comments by LLMs. 
Third, faced with the complex issue report, LLMs may generate hallucinations~\cite{2024arXiv240101313T}, producing irrelevant or unfactual comments. Reducing these hallucinations in the generated comments presents a significant challenge.

To address the preceding challenges, we propose \textbf{IsComment}, an issue-based retrieval and verification approach for generating supplementary code comments. First, we analyze supplementary code comments and issues in ten popular real-world projects, discovering five main types of supplementary information that issue reports can provide.
Second, we retrieve issue sentences containing these types of information to generate the corresponding supplementary code comments.  
Finally,  we reduce hallucinations in the code comments by filtering out those sentences that are irrelevant to the code or unverifiable based on the issue report.

To evaluate our approach, we construct a new dataset specially focused on supplementary code comments, and conduct experiments on it. 
Our experimental results indicate that: (1) Compared with existing work and LLMs, our approach can generate supplementary code comments effectively, increasing the coverage of manual supplementary comments from 35.8\% to 88.4\% for GPT-4o. (2) Our generated comments provide rich and useful information beyond existing comments, and hallucinations can be reduced through the verification processing. (3) Our approach is capable of generating supplementary code comments for all methods associated with relevant issues. Many of the comment information are essential for developers to understand tricky code.

In summary, this paper makes the following contributions:
\begin{itemize}
\item An approach named IsComment, which is the first to integrate issue retrieval and verification for LLMs to code comment generation, capable of generating rich, reliable and useful supplementary code comments for programming understanding.

\item An empirical study of code supplementary information in issue reports, along with a dataset of supplementary code comments, identifying five common types of useful information that issue reports can provide for  comment generation.

\item A comprehensive evaluation of the proposed approach, demonstrating its effectiveness in generating supplementary code comments, reducing hallucinations, and providing more useful information for both methods with and without manual code comments.
\end{itemize} 

The rest of the paper is organized as follows.
Section~\ref{sec:sup} provides a preliminary study on code supplementary information in issues. Section~\ref{sec:approach} presents our proposed approach.
Section~\ref{sec:experimental setup} outlines our experimental setup. Section~\ref{sec:experimental result} reports the experimental results. Section~\ref{sec:discussion} gives some discussion. 
Section~\ref{sec:relatedwork} reviews the related work.
Finally, 
Section~\ref{sec:conclusion} concludes this paper.

\section{Code Supplementary Information in Issue}

\label{sec:sup}
In this section, we introduce a preliminary study on what information can issue provide for code understanding and supplementary code comments generation.

\subsection{Data Collection}


\begin{table*}[t]
\caption{Statistics of the dataset Issuecom}
\label{tab:dataset}
\resizebox{\textwidth}{!}{
\begin{threeparttable}
    
  \centering
  \begin{tabular}{c|c|c|c|c|ccc|ccc|ccc|ccc}
    \hline
      \multirow{2}*{Project} & \multirow{2}*{\#Method}&\multirow{2}*{\#IMethod}&\multirow{2}*{\#ISMethod} & \#Manual & \multicolumn{3}{c|}{\#Code Line} & \multicolumn{3}{c|}{Issue Len} & \multicolumn{3}{c|}{\#Comment Sentence} & \multicolumn{3}{c}{Comment Sentence Len} \\
      \cline{6-17}
      && &&Validated&min&max&average&min&max&average&min&max&average&min&max&average
      \\
    \hline
    ambari &5123&3334 (65.1\%)&1228 (23.9\%)& 31 & 3 & 69 & 15.6 & 63 & 957 & 396.8 & 1 & 2 & 1.1 & 5 & 29 & 12.6\\
    camel &10741&6212 (57.8\%)&3379 (31.4\%) & 57 & 3 & 28 & 3.7 & 31 & 969 & 358.0 & 1 & 6 & 1.4 & 4 & 33 & 15.0\\
    derby&5653&4810 (85.1\%)& 1790 (31.6\%)& 64 &3 &160 &17.5 & 50 & 2821 & 1246.9 & 1 & 4 & 1.2 & 5 & 53 & 16.6 \\
    flink &9571&6452 (67.4\%)&3125 (32.6\%)& 18 & 3 & 22 & 5.8 & 40 & 2025 &258.8 & 1 & 4 & 1.2 & 5 & 44 & 18.2\\
    hadoop &12764&10519 (82.4\%)&4447 (34.8\%)& 83 & 3 & 126 & 16.4 & 87 & 2472 & 1056.6 & 1 & 5 & 1.2 & 4 & 48 & 11.5\\ 
    hbase &10396&8586 (82.6\%)& 3230 (31.0\%)& 35 & 3 & 45 & 14.5 & 66 & 2336 & 913.9 & 1 & 4 & 1.3 & 4 & 32 & 12.3\\
    jackrabbit&5935&4129 (69.6\%)&1574 (26.5\%) & 27 & 3 & 38 & 9.3 & 43 & 2813 & 538.2 & 1 & 3 & 1.3 & 7 & 35 & 15.5\\
    lucene &10352&8368 (80.8\%)&3792 (36.6\%)& 69 & 3 & 118 & 15.8 & 44 & 3085 & 977.1 & 1 & 5 & 1.2 & 4 & 36 & 13.5\\
    pdfbox &6003&4241 (70.6\%)& 1581 (26.3\%)& 25 & 4 & 67 & 16.3 & 52 & 1585 & 512.1 & 1 & 3 & 1.2 & 8 & 43 & 18.2\\
    wicket&4511&3156 (69.9\%)& 1061 (23.5\%) & 34 & 4 & 44 & 8.2 & 30 & 2321 & 502 & 1 & 3 & 1.3 & 4 & 53 &15.8\\
     \hline
     Overall &8104& 5980 (73.8\%)& 2520 (31.0\%)& 443 & 3 & 160 & 13.2 & 30 & 3085 &787.1 & 1 & 6 & 1.3 & 4 & 53 &14.4\\
     \hline
  \end{tabular}
   \begin{tablenotes}
      \item (IMethod: Methods having issue; ISMethod: Methods having issue and potential supplementary code comments (MESIA$>$3) )
    \end{tablenotes}
  \end{threeparttable}
  }
  \vspace{-3mm}
\end{table*}

To investigate the information that issue reports can provide for generating supplementary code comments, we first construct a  supplementary code comments dataset, named \textbf{Issuecom}, from
10 popular real-world projects in the Apache community~\cite{Apache}. In the dataset, each sample includes three parts: a code method, the method's supplementary code comment, and an associated issue report that contains the supplementary code information in the comment. The details of the data collection process are described below.

\subsubsection{\textbf{Mining Method-Comments Pairs}} 
We mine pairs of a method and its comments from the project repositories. We focus on method-level comments because they are the most needed types of comments~\cite{10.1145/3510003.3510152}. To avoid code-comment inconsistency~\cite{8813274}, we use the JGit tool~\cite{JGit} to parse the commit history of the project repositories and extract only method-comments pairs that are submitted in a single commit.

\subsubsection{\textbf{Filtering for Supplementary Code Comments}}
Not all the mined comments are supplementary code comments. In practice, generic comments~\cite{10.1145/3510003.3510152} that provide little supplementary information beyond code exist widely.
To filter supplementary code comments, we use the MESIA metric proposed by recent work~\cite{MESIA}. It measures the supplementary extent of code comments via information theory. We follow its recommendation to filter out comments with MESIA value lower than 3 and reserve the remaining comments as potential supplementary code comments.

\subsubsection{\textbf{Establishing Comment-Issue Link}} We establish the link between the comments and the issue reports by analyzing the commit message. 
A good commit message often has an issue ID, indicating that the method and its comments are committed to resolve the issue. Such an issue report will probably contain the code supplementary information desired in the method's comments.
Specifically, we design regex to extract the issue ID in commit messages to establish the comment-issue link, and leave out comments without an issue ID to ensure data quality.

        
  

\subsubsection{\textbf{Analyzing Issue-Comment Overlap}}
We analyze issue-comment overlap to further mine comments with code supplementary information verifiable in the issue report. 
Specifically, we compare each comment sentence with the issue sentences, and keep those comment sentences having over 70\% of words overlapping with one issue sentence. Through the overlapping analysis, we mine a considerable amount of candidate code comments whose code supplementary information is likely available and verifiable by the issue report.

\subsubsection{\textbf{Manual Validation}}
We employ three developers to validate the mined code comments. These three developers all have more than 5 years of programming experience. They are asked to inspect the mined comments and their related issue reports independently, and judge whether the comment contains code supplementary information that is available and verifiable in the issue report. We reserve only those comments that successfully pass all the three developers' judgements.

Table~\ref{tab:dataset}  shows the details of our \textbf{Issuecom} dataset. Here we can see that: 1) On average, 73.8\% methods in the projects have issue, but only 31.0\% of the methods have supplementary code comments. 2) The supplementary code comments can have multiple sentences, therefore can be longer than traditional summarizing code comments. 3) The issue report is much longer than the supplementary comment, and feeding the whole issue report to prompt LLMs will inevitably bring much noise and hallucination.

\subsection{Information Classification}

\begin{table}
\caption{Examples of different supplementary code comments}
\renewcommand{\arraystretch}{1.0}  
\setlength{\tabcolsep}{5pt}        
\label{tab:typeExample}
\centering
\small
\begin{tabular}{m{1.8cm}<{\centering} m{4.2cm} m{1.4cm}<{\centering}}  
\hline
\textbf{Method}                   & 
\centering \textbf{Comment}                                                                                                                  & \textbf{Type}          \\
\hline
toInputStream & Converts the given File with the given charset to InputStream with the JVM default charset.                                                  & Functionality \\
\hline
getZombieLeader & Zombie leader is a replica won the election but does not exist in cluster state.                                                           & Concept       \\
\hline
snapshot & Must be followed by a call to clearSnapshot(SortedMap)                                                                                     & Directive     \\
\hline
getPredicateBean & When sending messages to the control channel without using a DynamicRouterControlMessage, specify the Predicate by using this URI param. & Rationale     \\
\hline
setBulkRequests & Increasing this value may slightly improve file transfer speed but will increase memory usage.                                           & Implication  \\
\hline
\end{tabular}
\vspace{-5mm}
\end{table}

\begin{figure}[t]
    \centering
        \includegraphics[width=\linewidth,]{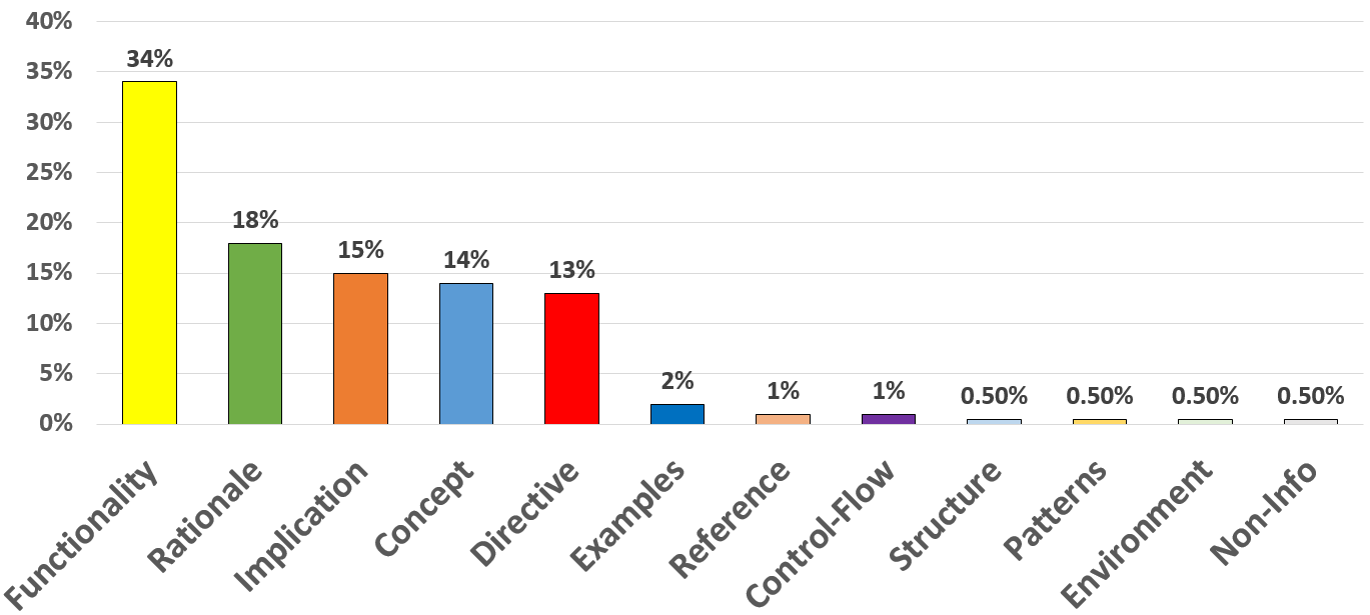}
    \vspace{-5mm}\caption{Proportion of different  supplementary code comments}
      \label{fig:Category}
      \vspace{-5mm}
\end{figure}

Based on the dataset, we employ three developers to manually classify the comments according to an existing taxonomy~\cite{6473801}. Compared with intention-based (What, Why, How, etc.) classification~\cite{10.1145/3434280,2023arXiv230207055M,2023arXiv230411384G}, this taxonomy provides a more concrete classification of code comments and can help developers better understand the supplementary information of the code. Specifically, the taxonomy originally contains 12 code comment categories, including Functionality, Concept, Directive, Rationale, Quality, Control-Flow, Structure, Patterns, Examples, Environment, Reference, and Non-info.
Table~\ref{tab:typeExample} shows some examples of classification. For example, developers classify the comment \emph{``Zombie leader is a replica won the election but does not exist in cluster state.''} into the concept category because the comment explains the concept of zombie leader. In the classification, developers perceive that comments in the "Quality" category describe a special quality about performance implications. To make it more specific, we rename this category as ``Implication" in our work.

We perform a two-round classification. In each round, three developers independently classify the comments according to the taxonomy. 
After the two-round classification, developers are required to discuss the potential disagreements and come to a consensus. 
Note that a comment may be classified into more than one category because it may contain multiple types of supplementary information.

Finally, we calculate the proportion of different categories of comments. The result is shown in Fig~\ref{fig:Category}.  We can see that most of the comments (94\%) fall into five main categories:

\begin{itemize}

\item \textbf{Functionality}, describing what the method does in terms of functionality, behavior, or feature.

\item \textbf{Rationale}, describing the purpose or rationale of the code, such as why the code is designed like this. 

\item \textbf{Implication}, describing the quality attribute with respect to the performance implications of the method.

\item \textbf{Concept}, explaining the meaning of the terms used in the method, especially some domain concept. 

\item \textbf{Directive}, describing what developers need to notice, such as what is allowed or not allowed, and some contracts.

\end{itemize}

Therefore, the rest of the paper focuses on retrieving the five common types of supplementary information from issue reports and generating the corresponding supplementary code comments.

\begin{result-rq}{Summary for Preliminary Study}
 The code supplementary information in the issue reports mainly includes Functionality, Rationale, Implication, Concept, and Directive. 
 These kinds of information are hard to derive by reading the source code.
\end{result-rq}

\begin{figure*}[t]
    \centering
      \includegraphics[width=1.0\linewidth,]{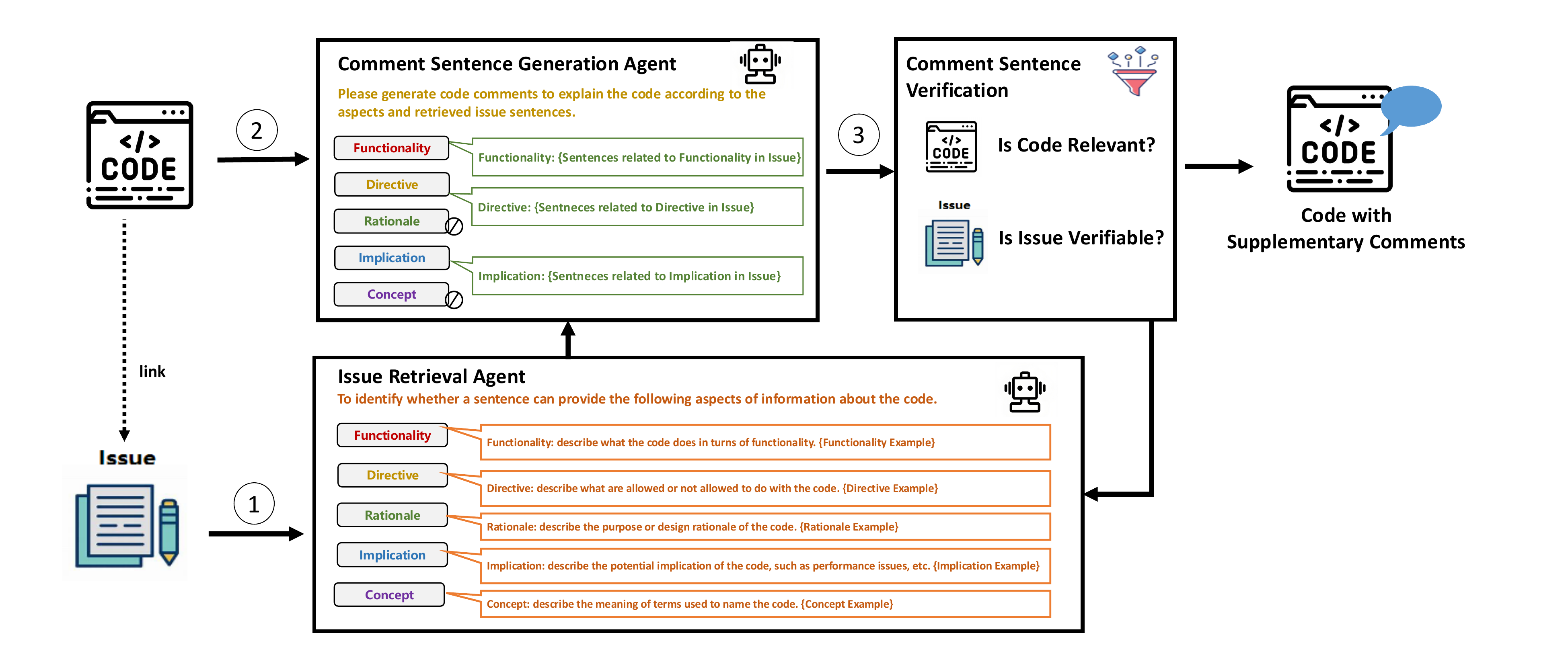}
      \vspace{-5mm}
    \caption{Framework of our approach.}
      \label{fig:framework}
      \vspace{-3mm}
\end{figure*}





\section{Comment Generation Approach}

\label{sec:approach}
In this section, we introduce our comment generation approach 
named \textbf{IsComment}. 
As shown in Fig~\ref{fig:framework}, it consists of three phases: issue retrieval, comment sentence generation, and comment sentence verification.


\subsection{Issue Retrieval}
In this phase, we aim to retrieve the issue sentences containing the potential five types of code supplementary information.

As an issue report is usually lengthy and miscellaneous, we first explored various common retrieval techniques, including TF-IDF~\cite{TF-IDF}, DPR~\cite{DPR}, and DistilRoBERTa~\cite{distilroberta}, to retrieve related issue sentences using the code method. 
However, we have observed that the retrieval results are not satisfactory because the retriever often returns irrelevant issue sentences or misses some types of desired code supplementary information.
Based on the above observation, we decide to take the code method and its associated issue report as input, prompting the powerful LLMs to verify and retrieve issue sentences containing the potential types of code supplementary information. The output is the types of code supplementary information and corresponding issue sentences for each type. Our prompt consists of three parts: (1) The description of the issue retrieval task; (2) The introduction to the five types of code supplementary information desirable by the target code method; (3) The target code method to be commented on and its associated issue report. The details of the prompt are available in our Appendix~\cite{Appendix}.

\subsection{Comment Sentence Generation}
In this phase, we aim to generate the supplementary code comments based on the retrieved issue sentences in each type.

We have observed that the retrieved issue sentences can be duplicated, disorganized, and incoherent, making them unsuitable for direct use as comments. 
As a result, we take the code method and the extracted issue sentences in each type to prompt LLMs to generate the corresponding supplementary code comments. Based on the types of code supplementary information, the outputs can have up to five distinct types of supplementary code comments. 
Our prompt consists of three parts: 
(1) The description of the comment generation task; 
(2) The types of code supplementary information along with the corresponding retrieved issue sentences; 
(3) The code method to be commented on. The details of the prompt are also available in our Appendix~\cite{Appendix}.

\subsection{Comment Sentence Verification}
In this phase, we aim to filter out comment sentences irrelevant to the code or unverifiable by the issue report to reduce hallucinations.

We first use two criteria to filter out comment sentences that are irrelevant to the code. One is whether the comment mentions any code elements (i.e., variable names or API names) in the code method. The other is whether the comment has a positive SIDE score~\cite{mastropaolo2024evaluating}, which is a newly proposed metric that uses the contrastive learning technology to learn code-comment semantic alignment and can evaluate how well the comment is semantically relevant to the code. Specifically, if neither of the two criteria is satisfied, we perceive the comment as irrelevant to the code and filter it out.


We further filter out comment sentences that cannot be verified by the issue report. The verification is based on the SentenceBert  metric~\cite{2019arXiv190810084R} recommended by existing work~\cite{10.1145/3524610.3527909}, which can measure the semantic similarity between the generated comment sentence and the issue sentence. It generates embeddings for words in a sentence, mean-pools these word embeddings to generate a sentence embedding, and computes the cosine similarity of the sentence embedding for the generated comment sentence and the issue sentence.  The calculation of the SentenceBert Similarity is presented in the following formulas, where $S_i$ and $I_j$ represent the generated comment sentence and the issue sentence, respectively:
\begin{equation}
\text{Vector}(x) = \text{MeanPooling}(\text{SentenceBert}(x)) \end{equation}
\begin{equation}
\resizebox{.9\hsize}{!}{$
\text{SentenceBert-Similarity}({S_i}, {I_j}) = \frac{\text{Vector}({S_i}) \cdot \text{Vector}({I_j})}{\|\text{Vector}({S_i})\| \|\text{Vector}({I_j})\|}$}
\end{equation}

For the generated code comments, we retain only those sentences that have at least one issue sentence with a similarity score above 0.6. The threshold 0.6 is a trade-off obtained through our manual attempt, and our validation of 100 pairs of sentences demonstrates that it can achieve an accuracy of 86\%, which is highly effective in evaluating the verifiability of the generated comments.


\section{Experimental Setup}

\label{sec:experimental setup}
This section describes research questions, dataset, comparing approaches, and evaluation setup.

\subsection{Research Questions}

\textbf{RQ1. How well can our approach improve LLMs in generating the supplementary code comments?} 

There is a lot of work on automatic comment generation. Answering this question helps investigate how well our approach can improve the generation of the supplementary code comments.

\textbf{RQ2. How well can our approach reduce the hallucinations hidden in the results?}

Compared with traditional approaches, LLMs can generate rich comments, including hallucinations irrelevant to the code or unverifiable. Answering this question helps investigate how well our approach can reduce hallucinations in the comment generation process.

\textbf{RQ3. How well are the supplementary nature of our generated comments? }

Answering this question helps investigate how well the generated code comments can bring verifiable supplementary information for developers in program understanding.

\textbf{RQ4. How well is the practical applicability of our approach in real-world projects? }

Answering this question helps investigate how well our approach can help generate the supplementary code comments using issue in real-world  project, especially for methods that have  no manual supplementary comments.

\subsection{Dataset} 

We use the \textbf{Issuecom} dataset rather than other existing code-comment datasets to conduct experiments for two main reasons. First, existing study~\cite{MESIA} has shown that the supplementary extents of code comments in existing dataset vary a lot, and not all comments in existing datasets are supplementary code comments. Second, it is hard to generate the supplementary code comments without additional resources beyond code such as issue. Compared to existing code-comment datasets, Issuecom not only provides the code and comment, but also provides the related issue, which can help  generate and verify the supplementary code comments.

\subsection{Comparing Approaches}
We compare our approach with the following approaches. For LLM-based approaches, the details of the prompts are available in our appendix~\cite{Appendix}.

\subsubsection{Non-LLM Approaches}
\begin{itemize}
\item \textbf{NCS (Ahmad et al.).} Such an approach is proposed by Ahmad et al.~\cite{ahmad2020transformer} and it trains a Transformer model with pairs of code-comment data in the TL-CodeSum dataset~\cite{TL-CodeSum} to generate code comments.

\item \textbf{CodeBert (Feng et al.).} 
Such an approach is proposed by Feng et al.~\cite{feng-etal-2020-codebert} and it fine-tunes the CodeBert pre-trained model with pairs of code-comment data in CodeSearchNet dataset~\cite{CodeSearchNet} to generate code comments.

\item \textbf{CodeT5 (Wang et al.).} 
Such an approach is proposed by Wang et al.~\cite{wang-etal-2021-codet5} and it fine-tunes the CodeT5 pre-trained model with pairs of code-comment data in CodeSearchNet dataset~\cite{CodeSearchNet} to generate code comments.

\end{itemize}
\subsubsection{LLM-based Approaches}

\begin{itemize}

\item \textbf{Code Prompt (Sun et al.).} Such an approach is proposed by Sun et al.~\cite{2023arXiv230512865S}. It provides only the code and a basic instruction for LLMs to generate code comments. 
Comparing with this approach helps investigate how well our approach can improve the generation of supplementary code comments with the help of issue reports.

\item \textbf{PS-Fewshot (Ahmed et al.).} Such an approach is proposed by Ahmed et al.~\cite{10.1145/3551349.3559555} and it adds 10 code-comment examples from the same project to the prompt for LLMs to generate code comments via in-context learning. Comparing with this approach helps investigate how well issue can help improve the generation of supplementary code comments compared with project examples.

\item \textbf{Code-Issue Prompt (No Retrieval).} This approach provides the code and the entire issue report for LLMs to generate code comments. Comparing with this approach helps investigate how well our type knowledge about the code supplementary information helps to improve the generation of supplementary code comments. 

\item \textbf{Code-Issue Retrieval Prompt.} This approach uses common information retrieval techniques to retrieve five most relevant sentences about the code from the issue report to augment the LLM prompt. Specifically, we explore three retrievers, namely \textbf{TF-IDF}~\cite{TF-IDF}, \textbf{DPR}~\cite{DPR}, and \textbf{DistilRoBERTa}~\cite{distilroberta}.
Comparing with such approaches can better understand the effectiveness of our approach.

\end{itemize}

\begin{table*}[t]
  \renewcommand{\arraystretch}{0.9}
  \caption{Comment Generation Results and Coverage Evaluation Results }
    \label{tab:RQ1Results}
  \centering
  \resizebox{0.9\textwidth}{!}{
  \begin{tabular}{c|l|c|c|c|c|c|c|c|c|c|c}
    \hline
      \multirow{3}*{LLM} & \multirow{3}*{Approach} & \multicolumn{5}{c|}{Before Filtering} & \multicolumn{5}{c}{After Filtering} \\
      \cline{3-12}
      && \#Sents&Sent Len &\#Full-&\#Partial- &  Coverage & \#Sents&Sent Len &\#Full-&\#Partial- &  Coverage
      \\
      &&(avg) &(avg)& Cover &Cover&(Ratio)&(avg) &(avg)& Cover &Cover&(Ratio)\\
    \hline
   \multirow{3}*{/}  &NCS (Ahmad et al.)&1.1 &11&23 &10&7.0\%& 0.1&8.2&  16&7&5.1\%\\

&CodeBert (Feng et al.)& 1.0&6.0&118 &36&34.7\%&0.5 &6.0&83  &27&24.8\%\\

&CodeT5 (Wang et al.)&1.0 &8.4&130 &33&36.7\%& 0.5&8.5&  94&28&27.5\%\\
     \hline
\multirow{7}*{ChatGPT}&Code Prompt (Sun et al.)&1.0 &18.6& 111&38&33.6\%&0.6 &19.5& 93 &33&28.4\%\\
&PS-Fewshot (Ahmed et al.)&1.7 &13.9&173 &41&48.3\%& 0.9&14.1& 140 &34&39.2\%\\
&No Retrieval&8.2 &13.2&257 &37&66.3\%& 3.9&15.4& 230 & 39& 60.7\%\\
&TF-IDF Retrieval& 5.1&12.8& 219&49& 60.4\%& 2.6&15.3&195  &49&55.0\%\\
&DPR Retrieval&5.1 &12.9& 237& 42& 62.9\%& 2.7&15.5& 208 &44&56.8\%\\
&DistilRoBERTa Retrieval&7.1 &14.3&267 &52&72.0\%&3.6 &16.8& 243 &52& 66.5\%\\
&\textbf{IsComment}&8.5 &13.8&278 &42&\textbf{72.2\%}$\uparrow$& 6.1&14.8& 254  &41& \textbf{66.5\%}$\uparrow$\\
    \hline
\multirow{7}*{GPT-4o}&Code Prompt (Sun et al.)& 3.4 & 8.5& 116&43&35.8\%& 0.6&19.5& 93 &33&28.4\%\\
&PS-Fewshot (Ahmed et al.)&1.2 &9.0&129 &39& 37.9\%& 0.5&10.3& 107  &33&31.6\%\\
&No Retrieval& 7.5 &16.8& 266 & 35& 67.9\%&4 &17.2& 239 &36& 62.0\%\\
&TF-IDF Retrieval& 7.4 & 15.6& 236& 32& 60.4\%& 3.4 & 16.4& 207  & 33& 54.1\%\\
&DPR Retrieval& 7.4 &16.6& 260 & 49& 69.7\%& 3.7&17.7&229  &50& 62.9\%\\
&DistilRoBERTa Retrieval& 7.7& 16.1& 279& 47& 73.5\%& 4.2& 16.9& 258 & 49& 69.3\%\\
&\textbf{IsComment}& 12.6 & 15.1& 358 & 34& \textbf{88.4\%}$\uparrow$& 8.8 & 15.9& 335  & 37& \textbf{83.9\%}$\uparrow$\\
     
     \hline
  \end{tabular}
  }
\end{table*}

\subsection{Evaluation}
We perform comprehensive evaluations for the generated code comments as follows.


\subsubsection{\textbf{Coverage Evaluation}}
To answer RQ1, 
we use the \textbf{SentenceBert Similarity}~\cite{2019arXiv190810084R} recommended by existing work~\cite{10.1145/3524610.3527909} to evaluate the generated comments. 
In particular, we split the rich generated comments and the manual comments into sentences. We regard a generated comment sentence successfully covers a manual comment sentence if their SentenceBert Similarity exceeds 0.6.  As one method may have more than one manual comment sentences, we use the number of methods whose manual comments are full-covered, partial-covered, or not-covered as evaluation results.



\subsubsection{\textbf{Verifiability  Evaluation}}
To answer RQ2,  we evaluate how well the generated comments are both code-relevant and issue-verifiable in our comment sentence verification phase mentioned earlier. To see how effective our evaluation criteria are, we apply them to the manual supplementary comments. 
87.1\% of the manual comment sentences are both code-relevant and issue-verifiable. Only one sentence fails both criteria. The result indicates the effectiveness of our evaluation approach to some extent.


\subsubsection{\textbf{Supplementarity Evaluation}}
To answer RQ3, we evaluate the supplementary nature of the generated code comments using the MESIA metric proposed by existing work~\cite{MESIA}. A larger MESIA value means a higher extent of supplementary information the comment can provide. 
After filtering out comments that are irrelevant to the code or unverifiable, such a metric can reflect how well the comments can provide beneficial supplementary information for developers.


\subsection{Model and Parameters}
 We use the ChatGPT (i.e., gpt-3.5 turbo~\cite{gpt-3.5-turbo}) and GPT-4o~\cite{gpt-4o} API to conduct our experiments. We set the temperature to zero to reduce randomness and keep the default values for other parameters.

\section{Experimental Result}

\label{sec:experimental result}
In this section, we present our experimental results in generating supplementary code comments using issue reports.

\subsection{RQ1. Supplementary code comment generation results}

\begin{figure}[t]
    \centering\includegraphics[width=\linewidth,]{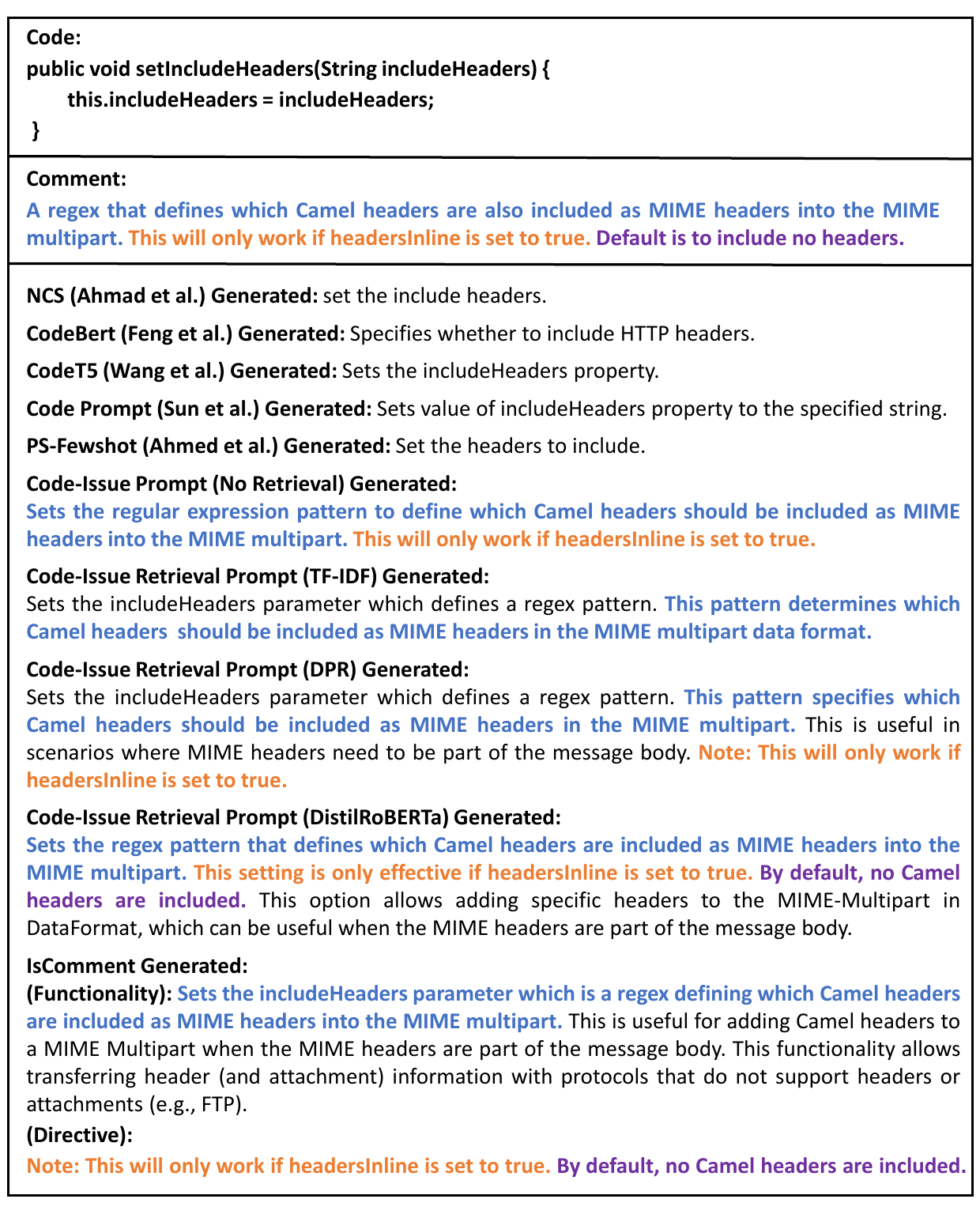}
      \vspace{-5mm}
    \caption{An example of code comments generated by different approaches.}
      \label{fig:RQ1Ex}
      \vspace{-5mm}
\end{figure}

\begin{figure*}[t]
    \centering
      \includegraphics[width=1.05\linewidth,]{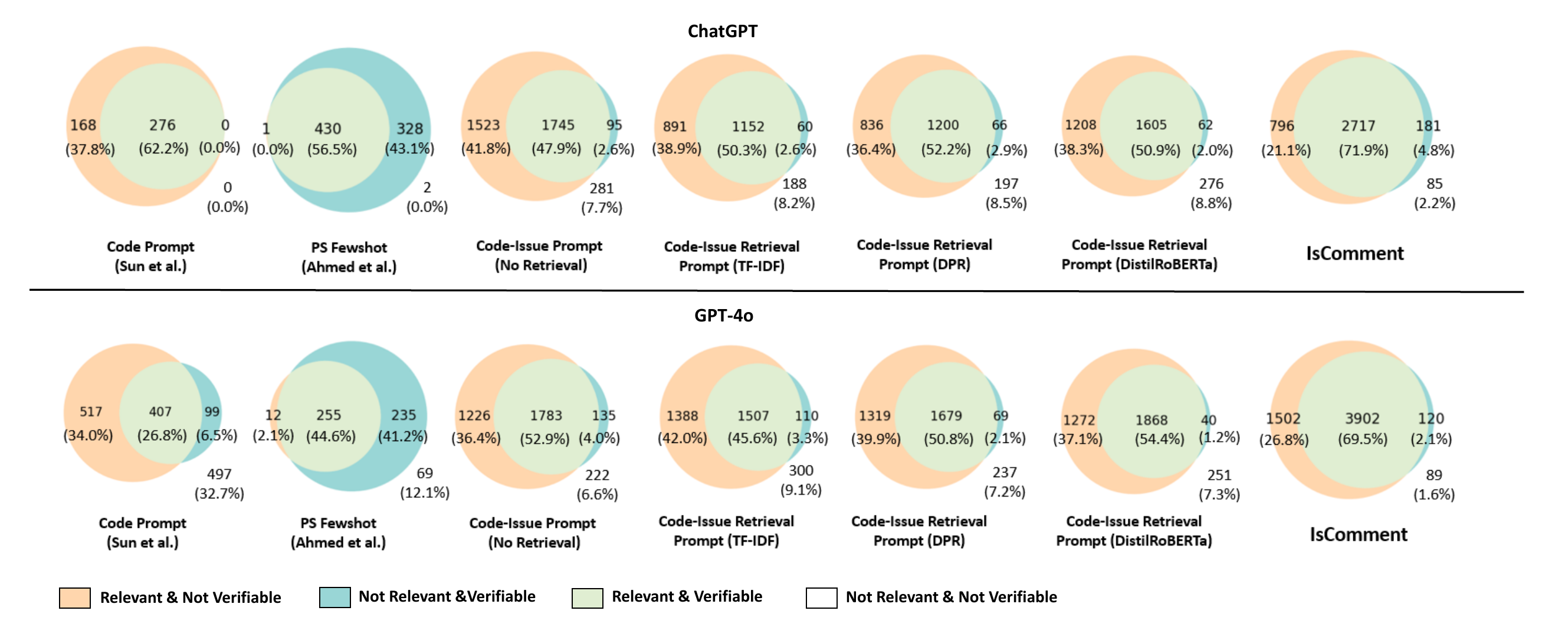}
         \vspace{-5mm}
    \caption{Evaluation results of the relevancy and verifiability of the comments generated by different approaches. }
    \vspace{-3mm}
      \label{fig:RQ2}
\end{figure*}


\begin{figure*}[t]
    \centering
      \includegraphics[width=0.87\linewidth,]{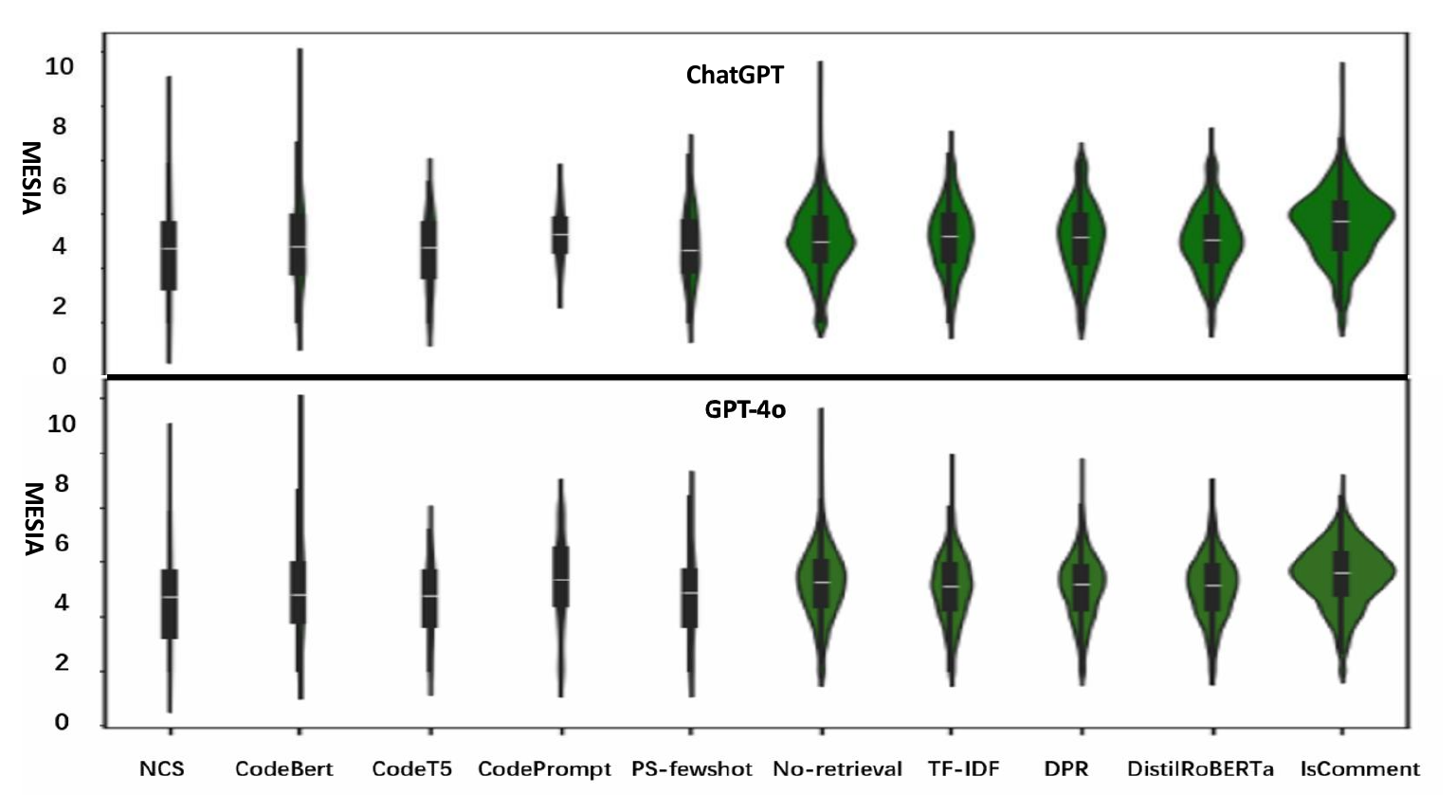}
      \vspace{-5mm}
    \caption{MESIA value of the  code comments generated by different approaches.}
      \label{fig:RQ3}
      \vspace{-4mm}
\end{figure*}

To answer this research question, we compare the coverage evaluation results among different approaches. 

Fig~\ref{fig:RQ1Ex} shows an example of the comments generated by different approaches using GPT-4o. 
For the method \emph{``setIncludeHeader"}, the manual supplementary comments have three sentences, including supplementary information about Funtionality and Dirictive about the method. We use different colors to distinguish these different comment sentences and use the same color to indicate the generated comments that cover the corresponding manual comment sentences. We can see that without issue, none of existing approaches can generate the supplementary code comments. As a comparison, our approach can successfully cover all the desired supplementary comment sentences.

Table~\ref{tab:RQ1Results} shows the overall results of the comments generated by different approaches. We can see from the table that
1) Traditional non-LLM approaches are difficult to generate supplementary code comments. After verification, none of the comment coverage in these approaches can reach 30\%. 
2) The issue is rather helpful for generating supplementary code comments for LLM-based approaches. For example, without issue, the Code Prompt approach in ChatGPT covers only 33.6\% of the manual supplementary comments and the coverage remains only 28.4\% after verification.
 But with the help of issue, the No-retrieval approach in ChatGPT can cover 66.3\% of the manual supplementary comments and the coverage still remains 60.7\% after verification. 
3) A traditional retriever does not necessarily improve the generation of supplementary code comments. For example, the generation results in two models are both worse than the No-retrieval approach when using the TF-IDF retriever, but the results can be improved when using the DistilRoBERTa retriever. This indicates that the retriever has a significant impact on the generation results.
4) Our approach IsComment can further improve LLMs to generate supplementary code comments. IsComment on GPT-4o can cover 88.4\% of manual code comments and coverage still remains 83.9\% after verification, significantly higher than other approaches, and the improvement is also higher than in ChatGPT. 
We suspect that this is because the retriever is a main bottleneck in our task, and more powerful LLMs can retrieve the code supplementary information better.

\begin{result-rq}{Summary for RQ1}
Our approach improves LLMs in generating supplementary code comments,  increasing the coverage of manual supplementary comments from 33.6\% to 72.2\% for ChatGPT and from 35.8\% to 88.4\% for GPT-4o. 
\end{result-rq}

\subsection{RQ2. Filtering out hallucination in the generated comments}


To answer this research question, we evaluated the code-relevancy and issue-verifiability of the generated comments in comment sentence verification phase. The results are presented in Fig~\ref{fig:RQ2}.

First, both code relevance and issue verifiability are helpful in reducing hallucinations. LLMs can generate many comments that although code-relevant but not issue-verifiable (e.g., 41.8\% comments of Code-Issue Prompt in ChatGPT), or issue-verifiable but not code-relevant (e.g., 43.1\% comments of PS-Fewshot in ChatGPT).
Considering both code-relevancy and issue-verifiability can ensure the reliability of the generated comments better.

Second, faced with the complex issue, our approach generates the fewest comments that are not code-relevant and not issue-verifiable. For example, for ChatGPT, Code-Issue prompt approach generates 281(7.7\%) comment sentences that are not code-relevant and not issue-verifiable, whereas our approach only generates 85 (2.2\%). We suspect that this is because our types of code supplementary information can better prompt LLMs.

Finally, after verification, the number of reliable comments in our approach is the best among all the approaches. In our approach, the proportion of comments that are code-relevant and issue-verifiable is 71.9\% for ChatGPT and is 69.5\% for GPT-4o, significantly higher than other approaches. After filtering, our approach generates 2717 sentences of comments in ChatGPT and 3902 sentences of comments in GPT-4o, significantly richer than other approaches. 




\begin{result-rq}{Summary for RQ2}
Both code-relevancy and issue-verifiability are helpful in reducing hallucinations of LLM-generated comments. The quantity of reliable comments generated by our approach is the best among all the approaches.

\end{result-rq}

\begin{figure*}[t]
    \centering
      \includegraphics[width=\linewidth,]{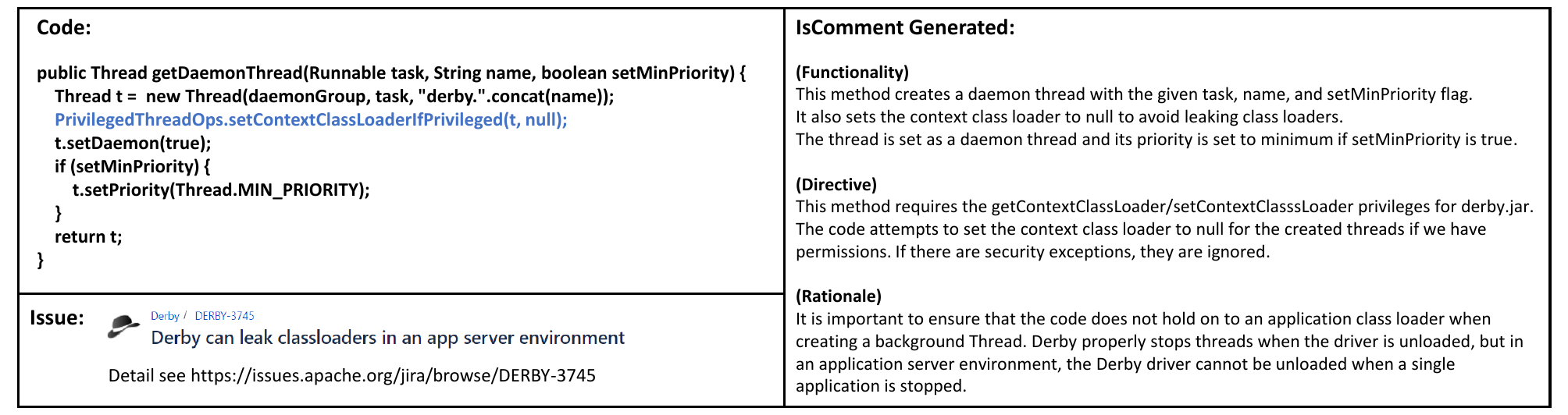}
      \vspace{-5mm}
    \caption{An example of code comments generated by our approach.}
      \label{fig:DiscussEx}
      \vspace{-4mm}
\end{figure*}

\subsection{RQ3. The supplementary nature of the generated comments} 

To answer this research question, we calculate the MESIA value for the verifiable comments of different approaches after filtering comments not code-relevant or not issue-verifiable. Fig~\ref{fig:RQ3} presents the results. In the figure, the higher altitude indicates a larger MESIA value and the larger area indicates a larger number of comments. From the figure,  we can understand the supplementary nature of the generated comments.

 First, our approach can generate more verifiable comments with a large extent of code supplementary information. For example, in GPT-4o, our approach generates 3902 comment sentences, which is 2.2 times the number of comment sentences generated by the No-retrieval method (1783). In addition, the comments generated by our approach have a better average MESIA value. For example, in ChatGPT, the average MESIA value of our generated comments is 4.57, larger than the average MESIA value (4.22) of comments generated by the No-retrieval approach. The results are similar for GPT-4o.

Second, our generated comments are rich and they can cover various  aspects about code. We provide the label of comment category in the generated comments so that they are in a more structured form. In practice, developers can choose their desired comments based on their own comment intentions. Because these supplementary comments have explicit evidence from issue, developers can accept them with more confidence.

\begin{result-rq}{Summary for RQ3}
Our approach can generate more verifiable supplementary comments than comparing approaches. These comments are more reliable with explicit evidence from the issue.
\end{result-rq}

\begin{figure}[t]
    \centering\includegraphics[width=\linewidth,]{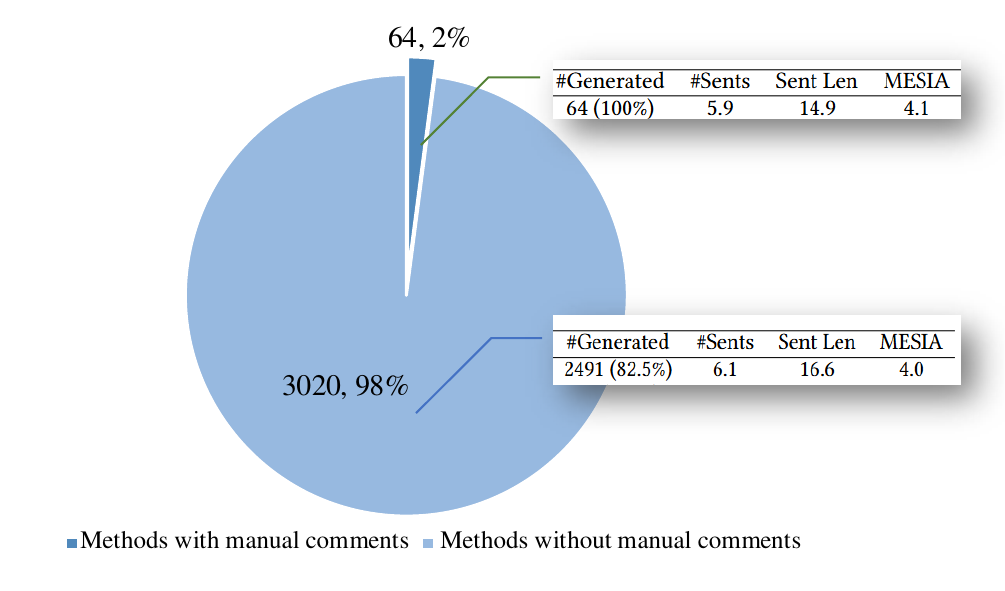}
      \vspace{-8mm}
    \caption{Supplementary comment generation results for methods in Apache Derby project.}
      \label{fig:DERBY}
      \vspace{-5mm}
\end{figure}

\subsection{RQ4. Application of Our Approach}

 In this section, we try to apply IsComment to generate supplementary code comments for some code (methods) that has related issue discussion but without manual comments. 

Due to time constraints, we now complete the experiments in one project of dataset, the Apache Derby project. As shown in Fig~\ref{fig:DERBY}, there are 3084 methods with associated issue reports, but only 64 methods have manual supplementary comments, while 3020 of them lack manual comments. Our approach successfully generates 2491 supplementary comments for  the methods without manual comments, achieving a generation rate of 82.5\%. Compared to the generation results for the 64  methods with manual code comments, which belong to our Issuecom dataset, the number of comment sentences, the sentence length, and the MESIA value of generated comments are very similar. 
This validates the practical applicability of our work.

Fig~\ref{fig:DiscussEx} shows a real-world example in the Apache Derby project. We notice that the tricky method \emph{getDaemonThread} requires special caution because it particularly invokes the \emph{setContextClassLoaderIfPrivileged} API to avoid classloaders leakage. For such a method without manual code comments, our approach can generate supplementary comments with rich information about its functionality, directive, and rationales. Many of the code supplementary information are never involved in previous work but they are especially essential for developers to understand such tricky code.

      

\begin{result-rq}{Summary for RQ4}
Our approach is capable of generating supplementary code comments for all methods associated with relevant issues. Many of the comment information are essential for developers to understand tricky code.
\end{result-rq}


\section{Discussion}

\label{sec:discussion}




In this section, we discuss the implication for future work and the threats to validity.

\subsection{Implication for Future Work}

\begin{table*}[t]
  \renewcommand{\arraystretch}{0.9}
  \caption{Comment Generation Results on DeepSeek-V3 }
    \label{tab:DeepSeek}
  \centering
  \resizebox{0.9\textwidth}{!}{
  \begin{tabular}{c|l|c|c|c|c|c|c|c|c|c|c}
    \hline
      \multirow{3}*{LLM} & \multirow{3}*{Approach} & \multicolumn{5}{c|}{Before Filtering} & \multicolumn{5}{c}{After Filtering} \\
      \cline{3-12}
      && \#Sents&Sent Len &\#Full-&\#Partial- &  Coverage & \#Sents&Sent Len &\#Full-&\#Partial- &  Coverage
      \\
      &&(avg) &(avg)& Cover &Cover&(Ratio)&(avg) &(avg)& Cover &Cover&(Ratio)\\
    \hline
\multirow{2}*{DeepSeek-V3}&Code Prompt (Sun et al.)& 1.0 & 23.5 & 113 & 42 & 35.0\% & 0.6 & 24.3 & 98 & 35 & 30.0\% \\
&\textbf{IsComment}& 6.7 & 20.2 & 332 & 50 & \textbf{86.2\%} $\uparrow$ & 4.7 & 20.7 & 310 & 49 & \textbf{81.0\%} $\uparrow$ \\
     
     \hline
  \end{tabular}
  }
    \vspace{-3mm}
\end{table*}


The rapid advancement of LLMs has led to numerous new models. One might assume that more powerful models would exhibit superior performance and notably reduce hallucinations.
However, our experiments in RQ2 show that hallucinations do not decrease even using a more powerful LLM. As our approach is adaptable and can be integrated with various LLMs, we also integrate it with DeepSeek-V3~\cite{deepseekai2024deepseekv3technicalreport}, which has attracted significant research attention recently. 
The results are shown in Table~\ref{tab:DeepSeek}, it also indicates that the number of comment sentences and the coverage of manual comments both decrease after filtering. 
We plan to conduct experiments with a wider range of LLMs to assess their performance more comprehensively.

Existing studies have reported the remarkable capability of LLMs in automated comment generation~\cite{10.1145/3551349.3559548,10.1145/3551349.3559555}, and LLM-generated comments can sometimes even surpass human-written ones~\cite{2023arXiv230909558P}. 
Our work indicates that this remarkable capability may result in an over-optimistic point. Our work conveys an important message that code comment generation still requires substantial research, whereas more research effort should be devoted to generate and evaluate supplementary code comments.

Our work points out that verification is necessary to generate code comments using LLMs. Traditional reference-based evaluation metrics ~\cite{10.1145/3510003.3510152,shi2022evaluation,10.1145/3524610.3527909}, such as the BLEU ~\cite{papineni2002bleu}, regards manual reference comments as the basis of the evaluation. However, LLM-generated comments may differ from reference comments, yet still be acceptable~\cite{mastropaolo2024evaluating,wu2024can}.
To address this, we split the generated comments and the manual reference comments into sentences and use a semantic similarity-based metric to evaluate coverage. 
Moreover, we perform code-relevancy and issue-verifiability evaluation to filter out the hallucinations. 
Our evaluation framework, including MESIA, can offer insight for better evaluating LLM-generated comments. 
In the future, we also plan to further integrate different code comment evaluation methods ~\cite{2023arXiv230402554G,wu2024can} into our evaluation framework.

Our approach could also adapt to other external resources. In practice, not all projects have high-quality issue reports, but other resources, such as requirement documentation, pull requests, or email discussions may also contain rich additional code information. Therefore, future work can explore applying our approach to these resources to generate supplementary code comments. 

\subsection{Threats to Validity} 
\label{sec:thread to validity}

The threats to validity include mainly the following three aspects.

The first is the dataset used in this paper. 
Since no existing dataset specifically focuses on supplementary code comments, we created a new dataset from scratch by mining data from 10 well-maintained popular projects across various domains.
We use strict criterion to extract the comments to ensure their quality. As a result, the size of the dataset is still small now.  We will expand our dataset in the future. Besides, we noticed that LLMs such as ChatGPT may have ``seen" the data of these projects, making the generation results of Code Prompt better than expected. However, our experiments indicate that without the issue report, they cannot generate the supplementary comments well.

The second is the prompts. Due to budget limitations, the prompts explored in this paper are currently limited. They may not fully unleash LLMs' capability in code comment generation, and there is still room for improvement. We will explore more effective prompts for LLMs to improve the generation of supplementary code comments in the future.

The third is our evaluation metrics. We discuss the concerns of traditional evaluation and suggest the necessity of verifiable generation. As a result, we conduct a comprehensive evaluation from the perspectives of coverage, verifiability, and supplementarity. However, more effective approaches and experiments are required in the future for a better evaluation.

\section{Related Work}

\label{sec:relatedwork}
In this section, we describe the related work, including (1) LLM-based code comment generation. (2) Issue report analysis and summarization. (3) Verifiable text generation.

\subsection{LLM-based Code Comment Generation}
LLMs have greatly promoted automatic code comment generation. 
For example, in 2022, Khan and Uddin~\cite{10.1145/3551349.3559548} first use Codex for this task and achieve significant better performance than traditional neural approaches. Later, Ahmed and Devanbu~\cite{10.1145/3551349.3559555} use Codex to generate code comments in project-specific scenarios and also achieve state-of-the-art. 
Lomshakov et al.~\cite{emnlp/LomshakovPSBLN24} propose to construct precise and efficient project-level contexts to enhance LLM-based code summarization. 
Geng et al.~\cite{10.1145/3597503.3608134} find that LLMs can generate code comments with various intentions. Ahmed et al.~\cite{10.1145/3597503.3639183} use code semantics to augment the LLM prompt to improve code comment generation. Sun et al.~\cite{2024arXiv240707959S} conduct a systematic and comprehensive study to investigate the capability of LLMs in code comment generation. Sun et al.~\cite{2023arXiv230512865S} report that the LLM ChatGPT can generate rich and fluent code comments. 

Existing LLM-based approaches still generate code comments based on the source code. As a result, they  struggle with generating supplementary code comments. 
our approach is the first to integrate issue retrieval and verification for LLMs to code comment generation, which can generate rich, reliable and useful supplementary code comments for programming understanding.

\subsection{Issue Report Analysis and Summarization}
Various approaches have been proposed to analyze or summarize the issue report. For example, Arya et al.~\cite{Arya2019AnalysisAD} conduct a qualitative analysis of the issue reports and automatically detect 16 types of information in issue reports. Rastkar et al.~\cite{6704866} explore using conversation-based automated summarizers to summarize the issue reports. Gilmer et al.~\cite{2023arXiv230802780G} propose techniques for issue users to collectively summarize different types of information in issue reports. Kumar et al.~\cite{10336257} explore the use of LLMs to summarize the issue report. 
In addition, Li et al.~\cite{9678864} generate issue ID in code comments to facilitate the understanding of code fragments using issue reports. Panichella et al.~\cite{6240510} use API method names to retrieve paragraphs of code descriptions from issue reports.
 

In summary, the preceding work focuses on analyzing or summarizing the issue report itself and
 do not pay attention to generating supplementary code comments. Our approach comprehensively analyzes the code supplementary information in issue reports and uses LLMs to generate supplementary code comments. 
\subsection{Verifiable Text Generation}
Verifiable text generation~\cite{li-etal-2024-llatrieval} aims to generate text content that has supporting evidence. Compared with Retrieval-Augmented Generation (RAG)~\cite{2023arXiv231210997G} that retrieves external resources to augment the LLM prompt, verifiable text generation further requires the generation results to be verifiable to ensure their reliability. In recent years, various studies in natural language processing have been devoted to verifiable text generation. For example, 
Li et al.~\cite{emnlp/SunCWHWWZY24} propose a framework with evolving memory and self-reflection for verifiable text generation.
Hennigen et al.~\cite{2023arXiv231109188T} propose symbolically grounded generation for easier validation of an LLM generation. Gao et al.~\cite{acl/Cao024} explore verifiable generation with subsentence-level fine-grained citations. Li et al.~\cite{li-etal-2024-llatrieval} explore using LLMs to verify the retrieving results for verifiable text generation.


In summary, the preceding work has achieved promising results in verifiable text generation. This paper applies this idea in software engineering and leverages issue reports for verifiable generation of supplementary code comments, which makes the generated comments have explicitly supporting evidence from issue.

\section{Conclusion}

\label{sec:conclusion}
Compared with existing code summarization approaches, our approach can leverage the issue to generate rich and helpful supplementary code comments for the methods.
Utilizing diverse LLMs, our approach places particular emphasis on issue retrieval and verification during the comment generation process, so as to effectively retrieve the related information from issue report and reduce hallucinations. 
Through our work, we envision a future where various forms of code supplementary information discussed in the software development process can be automatically identified, organized, and generated as code comments to facilitate program comprehension.

Based on our approach, we have implemented a prototype commenting tool in the form of a VSCode plugin. In future work, we plan to explore more effective ways to evaluate LLM-generated comments automatically. We also plan to explore more data sources beyond issue for code comments generation. 
Our code and data are available from: \url{https://github.com/Iscomment/IsComment}

\begin{acks}
This work is supported by the National Key Research and Development Program of China under Grant No.2023YFB4503803.
\end{acks}

\small
\balance

\bibliographystyle{unsrt}
\bibliography{reference}
\end{document}